\documentclass[11pt]{article}
\pdfoutput=1
\usepackage{jheppub}

\usepackage[usenames,dvipsnames,table]{xcolor}
\usepackage{graphicx,amsmath,amssymb,amsthm,multirow,array,bm,bbm,esint}
\usepackage[mathscr]{eucal}
\usepackage[bbgreekl]{mathbbol}
\usepackage{epsf,amsfonts}
\usepackage{slashed}
\usepackage[numbers,sort&compress]{natbib}
\usepackage{minibox}
\usepackage{array,tikz-cd}
\usepackage{slashed}
\usepackage{ccaption}

\definecolor{rust}{rgb}{0.8,0.2,0.2}

\newcommand{\prn}[1]{\left ( #1 \right )}

\newcommand{\bigbr}[1]{\Bigl\{ #1 \Bigr\} }

\newcommand{\half}{\frac{1}{2}}

\newcommand{\vev}[1]{\langle #1 \rangle}

\newcommand{\rhoi}{\hat{\rho}_{\text{initial}}}

\newcommand{\QSK}{\mathcal{Q}_{_{SK}}}
\newcommand{\QSKb}{\overline{\mathcal{Q}}_{_{SK}}}
\newcommand{\QKMS}{\mathcal{Q}_{_{KMS}}}
\newcommand{\QKMSb}{\overline{\mathcal{Q}}_{_{KMS}}}

\newcommand{\thetab}{\bar{\theta} }
\newcommand{\SF}[1]{\mathring{#1}}
\newcommand{\Dsf}{\SF{\mathcal D}}

\newcommand{\LamS}{\SF{\Lambda}}
\newcommand{\As}{\SF{\mathscr{A}}}
\newcommand{\Ascr}{\mathscr{A}}

\newcommand{\Fs}{\SF{\mathscr{F}}}
\newcommand{\Fscr}{\mathscr{F}}

\newcommand{\Kref}{{\bm \beta}}
\newcommand{\Lref}{\Lambda_\Kref}

\newcommand{\gref}{{\sf g}}
\newcommand{\Aref}{{\sf A}}
\newcommand{\shref}{{\sf h}}

\newcommand{\Lagref}{\mathscr L}
 \newcommand{\noiseg}{\mathfrak{N}}
 
\newcommand{\etaref}{\bm \eta}
\newcommand{\uref}{{\sf u}}
\newcommand{\Gref}{{\sf \mathbf G}}
\newcommand{\Nref}{{\sf \mathbf N}}
\makeatletter
  \newcommand\Ttiny{\@setfontsize\Ttiny{1pt}{2}}
\makeatother

\newcommand{\Tref}{{\sf T}}
\newcommand{\TEMref}{{\sf\mathbf T}}

\newcommand{\Pref}{{\sf P}}
\newcommand{\accref}{{\sf a}}
\newcommand{\sigref}{\sigma}
\newcommand{\omref}{\omega}
\newcommand{\Rref}{{\sf R}}
\newcommand{\Cref}{{\sf C}}

\newcommand{\lieD}{\pounds}

\newcommand{\Kbeta}{{\bm{\beta}}}
\newcommand{\LambdaB}{\Lambda_{\bm{\beta}}}

\newcommand{\Lag}{{\mathcal L}}

\newcommand{\skR}{\text{\tiny R}}
\newcommand{\skL}{\text{\tiny L}}

\newcommand{\smallT}{{\sf \!{\scriptscriptstyle{T}}}}

\newcommand{\UT}{U(1)_{\scriptstyle{\sf T}}}

\newcommand{\PS}{{\rm H}_S}
\newcommand{\PV}{{\rm H}_V}

\newcommand{\LS}{{\overline{\rm H}}_S}
\newcommand{\GV}{{\overline{\rm H}}_V}

\newcommand{\LT}{{\rm L}_{\,\smallT}}

\newcommand{\Wey}{{\scriptscriptstyle\mathcal{W}}}

\newcommand{\DWeyl}{\mathscr{D}^{\Wey}}

\title{Topological sigma models \& dissipative hydrodynamics}

\author[a]{Felix M. Haehl\textbf{}}
\author[b,c]{\!, R.\ Loganayagam}
\author[d]{\!, Mukund Rangamani}
\affiliation[\,a]{Centre for Particle Theory \& Department of Mathematical Sciences,\\
Durham University, South Road, Durham DH1 3LE, UK.}
\affiliation[\,b]{Institute for Advanced Study, Einstein Drive, Princeton, NJ 08540, USA.}
\affiliation[\,c]{ International Centre for Theoretical Sciences, Hesaraghatta Hobli, 
Bengaluru North,  560 089, India.}
\affiliation[\,d]{
Center for Quantum Mathematics and Physics (QMAP)  \\
Department of Physics, University of California, Davis, CA 95616 USA.}
\emailAdd{f.m.haehl@gmail.com}
\emailAdd{nayagam@gmail.com}
\emailAdd{mukund@physics.ucdavis.edu}

\abstract{
We outline a universal Schwinger-Keldysh effective theory which describes macroscopic thermal fluctuations of a relativistic field theory.  The basic ingredients of our construction are three: a doubling of degrees of freedom, an emergent abelian symmetry associated with entropy, and a  topological (BRST) supersymmetry imposing fluctuation-dissipation theorem.  We illustrate these ideas for a non-linear viscous fluid, and demonstrate that the resulting effective action obeys a generalized fluctuation-dissipation theorem, which guarantees a local form of the second law. 
}

\begin{document}
\maketitle

\section{Introduction}
\label{sec:intro}

Out of equilibrium quantum dynamics is relevant for a multitude of physical phenomena in many-body physics, quantum fields theories (QFTs), and in  quantum gravity. While the microscopic framework for studying this physics was pioneered in the 1960s by Schwinger, Keldysh  \cite{Schwinger:1960qe,Keldysh:1964ud} and others \cite{Feynman:1963fq}, a satisfactory theoretical derivation of the macrophysics remains elusive. 

Hydrodynamics, which is the universal low-energy physics of near-equilibrium thermal systems, exemplifies this status quo. Phenomenological aspects of fluid dynamics are well understood \cite{landau},  but a first principles derivation is lacking to date.  This is despite significant recent progress in understanding the essence of the hydrodynamics axioms \cite{Rangamani:2009xk,Hubeny:2011hd,Bhattacharyya:2012nq,Banerjee:2012iz,Jensen:2012jh,Bhattacharyya:2013lha,Bhattacharyya:2014bha,Haehl:2014zda,Haehl:2015pja} and attempts to write down a effective field theory for the same \cite{Nickel:2010pr,Dubovsky:2011sj,Dubovsky:2011sk, Endlich:2012vt,Haehl:2013hoa,Grozdanov:2013dba,Kovtun:2014hpa,Haehl:2015pja,Grozdanov:2015nea,Harder:2015nxa,Crossley:2015tka,deBoer:2015ija}. Such an effective theory is crucial to formulating a Wilsonian theory of (low-energy) fluctuations, complementing more phenomenologically oriented studies of macroscopic fluctuations based on large-deviation functionals \cite{Bertini:2015fk}.

Quantum gravity provides yet another motivation for seeking an effective low-energy formalism for mixed states. A SK-like doubling structure has been suggested to be crucial in understanding semiclassical physics of  AdS black hole interiors (and the experience of  an infalling observer) \cite{Papadodimas:2012aq}. Such proposals can be sensibly tested against  (better-understood) fluid dynamics  which, via AdS/CFT, is the correct nonlinear effective theory of the lowest quasinormal mode of the black hole \cite{Bhattacharyya:2008jc,Hubeny:2011hd}. We then need to understand in detail, not only how to embed fluid dynamics in the SK formalism at a non-linear level, but also how various ingredients in such a description get rephrased on the gravity side. This work should be viewed as a preliminary step towards such an endeavour.

In both black hole physics, as well as in the  fluid dynamics of the dual QFT, a major obstacle to writing down a unitary low-energy theory is the macroscopic non-unitarity and concomitant entropy production. This precludes a canonical effective field theory, for a type of UV-IR mixing below thermal energies obstructs the usual unitarity-preserving Wilsonian RG. AdS/CFT as usual geometrizes this: the scale/radius duality \cite{Susskind:1998dq} breaks down as one approaches the horizon (or probes sub-thermal scales in the QFT). This is also the crucial reason why the interior of the black hole cannot be given a simple interpretation via holographic renormalization. Any low-energy  theory should have a clear strategy to deal with this effective non-unitarity.

We recall that unitarity at the level of correlators is a twofold statement \cite{tHooft:1973pz}: (i) the correlators should factorize according to Cutkosky cutting rules (or more generally Veltman's largest time equation) and (ii) no ghosts should appear as the intermediate particles in the cutting rules. The properties of SK path integral  demand that the first structure be preserved in the low-energy effective theory, whereas the decoupling of ghosts may only happen after the effective theory is UV completed to a unitary theory. This idea is exemplified explicitly by Langevin theory in a Martin-Siggia-Rose \cite{Martin:1973zz} path integral description (for a review cf., \cite{ZinnJustin:2002ru}).\footnote{ We should clarify here that we are considering Langevin theory in a manifestly CPT invariant stochastic regularization, viz., Stratonovich regularization. In a simple theory like Langevin theory with an external (quenched) noise, one could decouple the ghosts by passing on to a  CPT breaking regularization (It\=o regularization) and then changing the rules of differentiation to It\=o calculus. While this is frequently done in order to avoid the ghosts, we would like to adhere here to standard QFT  with the usual rules of calculus.}  The cutting rule structure along with the fluctuation-dissipation relations and macroscopic second law heavily constrain the admissible effective non-unitarity  (see \cite{Haehl:2015foa} for details). Microscopically, these constraints originate from the Kubo-Martin-Schwinger(KMS) conditions which impose Euclidean periodicity on the SK correlators in the thermal limit.

The structural consequences of the second law in hydrodynamics are now well-understood  \cite{Bhattacharyya:2013lha,Bhattacharyya:2014bha,Haehl:2014zda,Haehl:2015pja}. This was used in  \cite{Haehl:2014zda,Haehl:2015pja,Haehl:2015foa}  to propose a general framework for an autonomous effective theory of hydrodynamics.  The macroscopic degrees of freedom were a thermal vector and twist (for flavour), $\{ \Kbeta^\mu  \equiv \frac{u^\mu}{T}, \ \LambdaB = \frac{\mu}{T} - \Kbeta^\alpha \,A_\alpha\}$  respectively, dynamics was the statement of current conservation (up to work terms), and a local form of the second law of thermodynamics was required to hold. A complete (eightfold) classification of hydrodynamic transport compatible with these axioms was derived in  \cite{Haehl:2014zda,Haehl:2015pja} and an effective action capturing seven of the eight classes (the adiabatic part) was constructed, aided by the observation of an emergent gauge symmetry, KMS invariance. The latter was postulated to be the macroscopic manifestation of the microscopic KMS conditions. 

These statements were further bolstered in \cite{Haehl:2015foa}, where we argued that the SK  construction and its concomitant KMS condition can be encapsulated in terms of a set of nilpotent BRST charges $\{\QSK,\QSKb,\QKMS,\QKMSb\}$, leading to an  ${\cal N}_T =2$ balanced topological theory. Such theories were first constructed in \cite{Vafa:1994tf} by topologically twisting ${\cal N}=4_\text{4d}$ SYM; the general formalism was described in  \cite{Dijkgraaf:1996tz}. We now  present a theory of dissipative hydrodynamics in this framework and realize our theory as a deformed, topologically twisted sigma model, with the hydrodynamic fields being the Goldstone modes for spontaneously broken difference diffeomorphism and flavour symmetries. For simplicity, we will only realize three of the eight classes (including dissipation) in the eightfold classification of \cite{Haehl:2015pja}. We also will demonstrate the validity of the second law, by deriving the generalized fluctuation-dissipation result of Jarzynski \cite{Jarzynski:1997aa,Jarzynski:1997ab} and Crooks \cite{Crooks:1998uq,Crooks:1999fk}, invoking spontaneous breaking of microscopic time-reversal as envisaged beautifully in \cite{Mallick:2010su,Gaspard:2012la,Gaspard:2013vl}. 
The construction we describe in the main text explicitly illustrates that the broad principles laid out in \cite{Haehl:2015foa} suffice to construct an effective field theory of dissipative hydrodynamics.

The rest of the paper is organized as follows: in \S\ref{sec:symsk} we outline the basic fields and symmetries, arguing that a superspace functional is the easiest route to our goal. We explain how these connect to the microscopic perspective in \S\ref{sec:fss} and proceed to exhibit an explicit construction for dissipative fluids in \S\ref{sec:nld}. We then demonstrate how to recover the generalized fluctuation-dissipation statement in \S\ref{sec:fd} and end with some comments in 
\S\ref{sec:discuss}.  We only sketch the basic principles here; full details of the construction will appear elsewhere \cite{Haehl:2015ab}.

\medskip\noindent
{\it Note:} Following \cite{Haehl:2015foa}, as this work was in progress, we received \cite{Crossley:2015evo} who also construct an  action for dissipative hydrodynamics based on principles of SK path integrals.

\section{Symmetries in SK description}
\label{sec:symsk}

We  begin by examining the fundamental symmetries of  a SK path integral. Given an initial density matrix $\rhoi$ of a QFT, we define the  SK generating functional
\begin{equation}\label{eq:SKdef}
 \mathcal{Z}_{SK}[J_\skR,J_\skL] \equiv \text{Tr}\bigbr{ U[J_\skR]\ \rhoi\  (U[J_\skL])^\dag } \,.
\end{equation} 
$U[J]$ denotes the unitary evolution of the QFT, deformed by a source $J$. This form of SK
functional immediately leads to a set of essential properties which should be satisfied by any SK effective theory \cite{Haehl:2015foa}.

\paragraph{Features for generic mixed states:}
First, when we align the sources  $J_\skR=J_\skL=J$, the SK functional localizes to $\rhoi$, viz.,
\begin{equation}
\mathcal{Z}_{SK}[J_\skR=J_\skL=J] \equiv \text{Tr}\bigbr{  \rhoi  }. 
\label{eq:eqsource}
\end{equation} 
This is a simple consequence of the unitarity of the underlying QFT. At the level of correlators, this 
implies that the  difference operators, $O_\skR-O_\skL$, form a protected topological subsector of the theory. This statement is equivalent  to the largest time equation/cutting 
rule for the corresponding correlator in the single copy theory. Thus imposing \eqref{eq:eqsource} in the 
low-energy effective theory ensures the cutting rule structure for its correlators.

This feature can be implemented in the SK effective theory by demanding that when sources align appropriately, the theory should exhibit topological invariance. Equivalently, any SK effective theory should be 
a \emph{source-deformed} topological theory (TQFT). Such a TQFT has two nilpotent, mutually anti-commuting,
Grassmann odd  topological charges $\{\QSK,\QSKb\}$, such that the difference operators are $\{\QSK,\QSKb\}$-exact  and SK effective action is $\{\QSK,\QSKb\}$-closed, modulo source terms proportional to $J_\skR-J_\skL$. 
When $J_\skR=J_\skL$, this theory naturally localizes as in \eqref{eq:eqsource}.

It is convenient to implement the topological invariance by working in superspace \cite{Horne:1988yn}. We introduce two Grassmann odd coordinates  $\{\theta,\thetab\}$, identify $\{\QSK,\QSKb\}\sim \{\partial_{\thetab},\partial_\theta\}$, and promote fields to superfields: 
\begin{equation}
{\cal Y} \to \SF{{\cal Y}} = {\cal Y} +  \theta\,  {\cal Y}_{\bar \psi} + \thetab \, {\cal Y}_\psi + \thetab\,\theta\, \tilde{{\cal Y}} \equiv 
\frac{{\cal Y}_\skL +{\cal Y}_\skR}{2} +  \theta\, {\cal Y}_{\bar \psi}  + \thetab \, {\cal Y}_\psi  + \thetab\,\theta\, ({\cal Y}_\skR -
{\cal Y}_\skL) \,.
\label{eq:xsfld}
\end{equation}	
The top ($\thetab\theta$) component of the superfields  represent the difference operators while the ${\cal Y}_\psi,
{\cal Y}_{\bar \psi} $ are the ghost super-partners of  ${\cal Y}$. Note that they carry the same spin as the field ${\cal Y}$ but opposite Grassmann parity. We can always recover the basic field by projection: 
\begin{equation}
{\cal Y} = \SF{{\cal Y}} | \equiv \SF{{\cal Y}} \big|_{\theta=\bar{\theta} =0} 
\,, \qquad 
\tilde{{\cal Y}} = \partial_\theta \partial_{\thetab} \SF{ {\cal Y}} | 
\equiv \partial_\theta \partial_{\thetab}\SF{{\cal Y}} \big|_{\theta=\bar{\theta} =0} \,.
\label{eq:snotation}
\end{equation}	
We will henceforth adhere to the convention that the circle $\mathring{}$ accent will denote the superfield corresponding to a field and tilde picks out the difference field in the SK construction.

Denoting spacetime 
coordinates by $\{\sigma^a\}$ and superspace coordinates by 
$z^I \equiv \{\sigma^a,\theta,\thetab\}$, we demand invariance under 
super-reparameterizations $z^I \mapsto f^I(z)$ for both (aligned) sources and fields.
Once  a TQFT has been constructed in the superspace, we can unalign the sources by
shifting the $\thetab\theta$ components of the sources thus breaking the topological invariance to get the 
required SK effective theory.

The second symmetry we implement is CPT, which implies that the SK path integral is invariant under
the combined  CPT transformation of the initial state and the sources. Using the anti-unitary 
nature of CPT, we can translate this into a reality condition for the SK path integral. It should
satisfy the identity (where $*$ represents complex conjugation)
\begin{equation}
\mathcal{Z}^*_{SK}[J_\skL,J_\skR] =  \mathcal{Z}_{SK}[J_\skR,J_\skL]\,.
\end{equation} 
This identity  follows simply from the definition in 
\eqref{eq:SKdef} along with hermiticity of $\rhoi$. 
As expected CPT acts anti-unitarily with a  complex conjugation; it exchanges the left and the right sources. Apart from the usual action on $\sigma^a$, CPT exchanges $\thetab\leftrightarrow\theta$ and hence acts as an R-parity on the superspace. This is necessitated by  our requirement that the $\thetab\theta$ component of the superfields be identified with difference operators.  It is further natural to extend these symmetries by including  ghost number conservation with  $\thetab$ and $\theta$ having opposite ghost numbers (wlog $\mp1$).  The super-reparametrization invariance, CPT invariance and ghost number conservation form the basic set of symmetries to be imposed on any SK effective theory.\footnote{ Thus in particular, these structures should also be present when we consider reduced density matrices for some spatial region of a QFT as is usually done in the context of entanglement entropy.}

\paragraph{Features of near-thermal density matrices:}
We now move on to symmetries specific to near-equilibrium situations. For thermal $\rhoi$, SK correlators can be obtained by analytically continuing Euclidean thermal correlators. Under this continuation, Euclidean thermal periodicity translates into a set of non-local KMS conditions \cite{Kubo:1957mj,Martin:1959jp,Haag:1967sg}.  They characterize the UV-IR mixing inherent in thermal states, with the scale of non-locality being the thermal scale. Any effective theory of near-equilibrium fluctuations should efficiently encode these conditions non-linearly. This problem is well-studied  (but without clear resolution) for non-relativistic systems in macroscopic fluctuation \cite{Bertini:2015fk} and mode-coupling theories. The issue is one of implementing fluctuation-dissipation relations at the non-linear level. One may of course impose the KMS relations directly on the correlators by hand, but it is unclear how to maintain them under renormalization.

Inspired by our previous studies of the structural consequences of the second law in relativistic fluids, we had advocated a solution to this  conundrum in terms of an emergent $\UT$ gauge invariance \cite{Haehl:2014zda,Haehl:2015pja,Haehl:2015foa}. This KMS symmetry acts on the fields by thermal translations. In particular:
\begin{enumerate}
\item[(a).] It ensures the correct localization of the SK path integral satisfying Euclidean periodicity, by extending the cohomology of   $\{\QSK,\QSKb\}$ into an equivariant cohomology of thermal translations. 
\item[(b).]  It gives rise to a macroscopic entropy current thus intimately linking the emergence of entropy with the microscopic KMS invariance. 
\end{enumerate}
In the gravitational description, this statement is then dual to Wald's construction of black hole entropy as a Noether charge \cite{Wald:1993nt,Iyer:1994ys}.

To describe our macroscopic gauge theory at a certain temperature we introduce a background timelike superfield $\Kref^a(\sigma)$. It can be viewed as a vector superfield  $\SF{\Kref}^I(z)$ with $\SF{\Kref}^\theta=\SF{\Kref}^{\thetab} = 0=\partial_\theta \SF{\Kref}^a=  \partial_{\thetab} \SF{\Kref}^a$. We will consider below only that subset of superdiffeomorphisms which respect this gauge choice for the background thermal supervector $\SF{\Kref}^I$. Similarly for charged fluids we introduce a  thermal twist $\Lref(\sigma)$ which encodes the chemical potential.\footnote{ This is the phase entering the thermal periodicity conditions in a particular flavour symmetry gauge.} These background fields  play a fundamental role in the gauge theory describing thermal fluctuations. 

The \emph{supergauge $\UT$ transformations} are parameterized by an adjoint superfield $\LamS$. They  
act on a general superfield $\SF{{\cal Y}}$ by Lie dragging it along $\LamS\, \Kref^a$. Such transformations can be succinctly represented by introducing a special type of Lie bracket which we christen as a \emph{thermal bracket},
\begin{equation}
(\LamS,\SF{{\cal Y}})_\Kref=\LamS  \,\lieD_\Kref \SF{{\cal Y}} \,,
\label{eq:betabrkX}
\end{equation}	
where  $\lieD_\Kref$ denotes the Lie-derivative along $\Kref^a$. The 
infinitesimal gauge transformation is thus given by
\begin{equation}
\SF{{\cal Y}}\mapsto \SF{{\cal Y}}+ (\LamS,\SF{{\cal Y}})_\Kref \,.
\label{eq:utx}
\end{equation}	
For scalar $\SF{{\cal Y}}_\text{scalar}$ this is  just a thermal translation\footnote{ If the superfield is also charged under flavour, this generalizes to $(\LamS,\SF{{\cal Y}})_\Kref=\LamS \prn{ \lieD_\Kref \SF{{\cal Y}}-[\Lref,\SF{{\cal Y}}]}$ where $[\,\cdot\,,\,\cdot\,]$
denotes flavour adjoint action.}  
\begin{align}
(\LamS,\SF{{\cal Y}}_\text{scalar})_\Kref=\LamS \;\Kref^a \partial_a \SF{{\cal Y}}_\text{scalar}\,.
\label{}
\end{align}
The Jacobi identity  then fixes the action of  thermal bracket on adjoint superfields, so that under $\UT$ transformation $\LamS' \mapsto \LamS' + (\LamS,\LamS')_\Kref $  with
\begin{align}
(\LamS,\LamS')_\Kref &=\LamS\lieD_\Kref \LamS'-\LamS' \lieD_\Kref \LamS \,.
\label{eq:adbetabrk}
\end{align}

We introduce  \emph{a gauge superfield one-form} as a triplet
 $\As_I(z)\equiv\{\As_a(z),\SF{\mathscr{A}}_{\theta}(z),\SF{\mathscr{A}}_{\thetab}(z)\}$, i.e.,
\begin{align}
\As_I(z)\, dz^I = \As_a(z)\, d\sigma^a +\SF{\mathscr{A}}_{\theta}(z)\, d\theta + \SF{\mathscr{A}}_{\thetab}(z) \, d\bar{\theta}
\label{eq:Aform}
\end{align}
  whose gauge transformation is like an adjoint superfield except for an inhomogeneous term, viz., 
\begin{equation}
 \As_I \mapsto \As_I +(\LamS,\As_I)_\Kref  - \partial_I \LamS \,,
\label{eq:aut}
\end{equation}	
with the thermal bracket as in \eqref{eq:adbetabrk}.  One can further define as usual a 
\emph{covariant derivative} 
\begin{equation}
 \Dsf_I = \partial_I + (\As_I,\ \cdot\, )_\Kref\,,
\label{eq:covDI}
\end{equation}	
and an associated field strength
\begin{equation}
\Fs_{IJ}  \equiv(1-\frac{1}{2}\, \delta_{IJ}) \left(\partial_I\, \As_J - (-)^{IJ} \,\partial_J\, \As_I  + (\As_I,\As_J)_\Kref \right) ,
\label{eq:fdef}
\end{equation}	
where $(-)^{IJ}$ is the mutual Grassmann parity of the two indices involved.
Given the low-energy superfields $\SF{{\cal Y}}$, the theory of macroscopic fluctuations is given as the general superspace action invariant under  $\UT$ gauge transformations. We have sketched in \cite{Haehl:2015foa} how this construction works for Langevin dynamics -- the full symmetry algebra can be found in Appendix A therein. 

In what follows we will suppress the details of the gauge sector for the most part, quoting only the key statements we need to write down the results. Thus, in our work the gauge sector will only appear as minimally coupled and we will systematically ignore the non-minimal
couplings and the detailed dynamics of the gauge sector in this work.
 A complete story involves explaining how the equivariant construction of $\UT$ dynamics works and will appear soon in our companion paper \cite{Haehl:2015ab}.\footnote{ The astute reader will also recognize that the  $\UT$ gauge symmetry despite being abelian has to act non-linearly, owing to its origin in thermal translations. Indeed the $\UT$ gauge theory has many features of a non-commutative abelian gauge theory. Heuristically this  may be motivated from our earlier statements about UV/IR mixing below the thermal scale.}

\section{Fields and symmetries}
\label{sec:fss}

Thus far we have argued that the symmetries needed to describe any theory of thermal fluctuations are:
\begin{enumerate}
\item Superdiffeomorphisms
\item CPT
\item Ghost number conservation
\item $\UT$ gauge invariance
\end{enumerate}
We now would like to implement these to study fluctuations in fluid dynamics. 

We start with the intuition that degrees of freedom are Goldstone modes for spontaneously broken off-diagonal diffeomorphism (and flavour) in the SK description \cite{Haehl:2015pja} (see also \cite{Nickel:2010pr,Kovtun:2014hpa}). The relevant Landau-Ginzburg sigma model is that of  a space-filling Brownian brane introduced in \cite{Haehl:2015foa}. These are  brane like objects of various codimension which, when immersed in the fluid, undergo generalized Brownian motion.

Focusing on  energy-momentum transport in neutral fluids,  the pion fields of the sigma model are vectors 
$X^\mu(\sigma)$ corresponding to the broken difference diffeomorphisms of the SK construction.\footnote{ The analogy with chiral symmetry breaking is quite apposite here. Note that the symmetry being broken is `universal' for any QFT -- thus our discussion at the structural level is agnostic of the actual QFT dynamics, which is what we expect for hydrodynamics. The details of the QFT start to matter if we actually ask for the precise values of the hydrodynamic transport data, which are the analog the pion coupling constants in the effective field theory.} They map the points in the worldvolume of the brane, parametrized by $\sigma^a$, to the physical target spacetime where the fluid lives. The fluid velocity and temperature are then described by a push-forward of the worldvolume $\Kref^a$ to the fluid dynamical spacetime using   $X^\mu(\sigma)$. Further, we obtain a worldvolume metric  $\gref_{ab}$  by pulling back the source $g_{\mu\nu}$ from the physical spacetime.\footnote{ A clarification is in order here. We constructed a  Landau-Ginzburg sigma model for adiabatic hydrodynamics in \cite{Haehl:2015pja}. There the physical fields were argued to be maps from physical spacetime onto a reference manifold (whose coordinates etc., were denoted by blackboard bold characters), cf., Figs. 3 and 4 therein. We call the reference manifold  as the worldvolume here to make the analogy with D-branes more explicit. This change of perspective makes generalizations to Brownian branes simpler. To keep the notation from getting out of hand, we have consequentially swapped the blackboard bold font used for reference fields in \cite{Haehl:2015pja}, for a simpler sans serif version here.}

 Our proposal is to take this  set-up and make it consistent with the symmetries given in \S\ref{sec:symsk}. This can be achieved by enhancing the  pull-back fields $X^\mu(\sigma)$ to superfields. Not only do the bosonic hydrodynamic pions get upgraded to a superfield, but we also obtain their Grassmann odd counterparts. This leads to a triplet of spacetime superfields    $\{\SF{X}^\mu(z), \SF{\Theta}(z),\SF{\bar{\Theta}}(z)\}$ on which super-diffeomorphisms, CPT, and ghost number symmetries act naturally. The action of $\UT$ is given as:
\begin{equation}
(\LamS,\SF{X}^\mu)_\Kref = \LamS \, \Kref^a \partial_a \SF{X}^\mu\,, \qquad
 \label{eq:xut}
\end{equation}
and similarly for $\{\SF{\Theta}(z),\SF{\bar{\Theta}}(z)\}$.
The worldvolume metric $\gref_{ab}$ gets upgraded to superfields
$\SF{\gref}_{IJ}$ using the $\UT$ covariant $\Dsf_I \SF{X}^\mu$:
\begin{align}\label{eq:gref}
\SF{\gref}_{IJ}(z) = g_{\mu\nu}(\SF{X}(z))\,\Dsf_I \SF{X}^\mu\,\Dsf_J \SF{X}^\nu \,.
\end{align}
Deformations away from the topological limit involve dealigning the sources for the left and right fields; for the energy-momentum tensor this can be achieved by turning on a difference source $\color{blue}{\shref_{IJ}}$, i.e.,
\begin{align}
\SF{\gref}_{IJ}(z) \;\rightarrow\; \SF{\gref}_{IJ}(z) + \color{blue}{\bar{\theta} \,\theta\;\shref_{IJ}(\sigma)}\,.
\label{eq:grefh}
\end{align}
Once we have an appropriate superspace Lagrangian, varying it with respect to the source deformation $\shref_{ab}$ will give us the (worldvolume) fluid dynamical stress tensor $\TEMref^{ab}_\text{wv}$, which can subsequently be pushed-forward to the physical target space to get $T^{\mu\nu}$.

A natural consequence of enhancing the target space fields to superfields is that the target space diffeomorphisms, CPT and flavour symmetry get enhanced to 
\begin{enumerate}
\item[A.] Target space super-diffeomorphisms of  $\{\SF{X}^\mu,
\SF{\Theta},\SF{\bar{\Theta}} \}$.
\item[B.] Target space CPT acting on $\{\SF{X}^\mu,
\SF{\Theta},\SF{\bar{\Theta}}\}$.
\end{enumerate}
These two symmetries, particular to fluid dynamics, along with the four symmetries enumerated above constitute
the complete set of symmetries to describe the macroscopic thermal fluctuations in fluid dynamics. In what follows, we will exploit a part of the target space super-diffeomorphisms to set $\{\SF{\Theta}=\theta,\SF{\bar{\Theta}}=\thetab\}$. The reader may find the analogy with the superstring worldsheet theory useful. The picture we portray above is the Ramond-Neveu-Schwarz formalism for the space filling Brownian brane. As discussed in \cite{Haehl:2015foa} the worldvolume TQFT can be similarly constructed for higher codimension Brownian branes, with the Brownian particle (or zero brane) theory leading to a description of Langevin dynamics.

We simply note in passing that  the above discussion can be extended to include other conserved charges. For flavour symmetry with source $A_\mu$ in the physical spacetime, the Goldstone modes include a flavour group element $c(\sigma)$, which map points on the flavour bundle of the worldvolume onto the physical flavour bundle. The chemical potential is defined by pushing-forward the worldvolume thermal twist $\Lref$. Moreover,  incorporating the desired supertransformations  one upgrades $c(\sigma)$ to a superfield $\SF{c}(z)$. 
This gives  a worldvolume  pull-back flavour  gauge field $\SF{\Aref}_a$ defined by the map $\{\SF{X}^\mu, \SF{c} \}$ which may further be deformed by dealigning sources (i.e., introduce $\alpha_a(\sigma)$).
The crucial item to note is the  $\UT$ transformation on the flavour superfield, which is given by
$$(\LamS,\SF{c})_\Kref = \LamS \, \SF{c}\, \prn{\Lref + \Kref^a \SF{\Aref}_a}\,. $$ 
 Finally, we should append to the list of symmetries A and B,  the target space flavour symmetry acting on $\SF{c}$.

\section{Non-linear dissipative fluids}
\label{sec:nld}

With the basic fields and symmetries in place, we are now in a position to construct an effective action. The symmetries 1-4 can be manifestly implemented by working in superspace. We then have focus on the target space symmetries A and B of \S\ref{sec:fss}. 

We begin by noting that the target space diffeomorphism invariance forbids a standard superpotential term, i.e., a function of the fields $\SF{X}^\mu$.   Consequentially, the simplest allowed term is a worldvolume scalar density superpotential, $\SF{\Lagref}\left[\SF{\gref}_{ab}, \Kref^a\right]$, which is a functional of the pull-back metric superfield $\SF{\gref}_{IJ}$. Such  terms (up to on-shell equivalence) comprise the Landau-Ginzburg Class L ($\PS \cup \LS$) in the classification of \cite{Haehl:2015pja}. They however are adiabatic and do not capture dissipative dynamics. 

To see dissipation, consider the superfields $\Dsf_\theta \SF{\gref}_{ab}$ and $\Dsf_{\thetab} \SF{\gref}_{cd}$ which carry non-zero (and opposite) ghost number. While neither of them can individually appear in the Lagrangian given our symmetries 1-4, we can combine them with an intertwining tensor, $\SF{\etaref}^{(ab)(cd)}$, of vanishing ghost number. In general this intertwiner may be taken to be a tensor valued differential operator, $\SF{\etaref}^{(ab)(cd)}[\SF{\gref}_{ab},\Kref^a,\Dsf_I]$ as in \cite{Haehl:2015pja} but we will focus on simple examples where it will suffice to think of it as a worldvolume tensor superfield. 

We therefore claim that the following worldvolume superspace action functional captures dissipative hydrodynamic effective field theories:
\begin{align}
 S_\text{wv} \equiv \int d^d\sigma\, \Lag_\text{wv}  \,, \qquad \Lag_\text{wv}= \int\,d\theta\, d\thetab\, \frac{\sqrt{-\SF{\gref}}}{1+\Kref^e \As_e} 
\left( \SF{\Lagref} - \frac{i}{4}\, \SF{\etaref}^{(ab)(cd)}\, \Dsf_\theta \SF{\gref}_{ab} \; \Dsf_{\thetab} \SF{\gref}_{cd} \right) \,,
\label{eq:petalag}
\end{align}	
where the measure is dictated by $\UT$ invariance.\footnote{ Note that the extra factor in the measure is just what is expected when working with a $\UT$ covariant pulled-back metric, for
$$ d^d\sigma \frac{\sqrt{ -\SF{\gref} } }{1+\Kref^a \As_a} =   d^d\sigma \sqrt{-\SF{\gref} }\, \frac{\text{det
}[\partial_a \SF{X}^\mu] }{\text{det} \; [\Dsf_a \SF{X}^\mu] } = d^d\SF{X} \sqrt{-g}\,.$$ }
CPT invariance  forces the tensor $\etaref^{(ab)(cd)}$ to satisfy the generalized Onsager reciprocity relations \cite{Onsager:1931fk,Onsager:1931uq}:
\begin{equation}
\SF{\etaref}^{(ab)(cd)}  = [ \SF{\etaref}^{(cd)(ab)} ]^{\text{\sf{\tiny CPT}}}\,,
\label{eq:etacpt}
\end{equation} 
where the superscript ${\sf {\tiny CPT}}$ on the right hand side denotes taking the CPT conjugate.

Let us first recover the familiar form of the hydrodynamic constitutive relations. To this end, we begin by defining 
\begin{equation}
\begin{split}
\TEMref_{\Lagref}^{ab} &\equiv \frac{2}{\sqrt{-\gref}} \, \frac{\delta }{\delta \gref_{ab}}[\sqrt{-\gref}\ \Lagref]\,,
\\ 
\Nref_{\Lagref}^a &\equiv -\frac{(1+\Kref^e \Ascr_e)}{\sqrt{-\gref}} \, \frac{\delta }{\delta \Ascr_{a}}\big{[}\frac{\sqrt{-\gref}}{(1+\Kref^f \Ascr_f)}\ \Lagref\big{]}\,, 
\end{split}
\label{}
\end{equation}	
where $\Lag \equiv \SF{\Lag}|$ and we treat $\{\Kref^a,\SF{\gref}_{IJ}, \As_I\}$ as independent fields for the purposes of functional differentiation. Performing the superspace integral in \eqref{eq:petalag} in a convenient gauge, we obtain
\begin{align}
\Lag_\text{wv} &=  \frac{\sqrt{-\gref}}{1+\Kref^e \Ascr_e} \;\bigg\{
	\frac{1}{2}\left[\TEMref_{\Lagref}^{ab}  - \frac{i}{2} \etaref^{(ab)(cd)}\, \left(  {\mathscr F}_{\theta\bar{\theta}}, 
	\gref_{cd}\right)_{\Kref}\right] \tilde{\gref}_{ab} 
\bigg. \nonumber \\
 & \qquad \qquad \qquad \bigg.
	 + \frac{i}{8}\, \left(\etaref^{(ab)(cd)} + \etaref^{(cd)(ab)} \right) \tilde{\gref}_{ab} \, \tilde{\gref}_{cd}
	-\Nref_{\Lagref}^a \tilde{\Ascr}_a+ \ldots \bigg\}\,,
\label{eq:classLTa}
\end{align}
where we have elided over the terms involving bilinears in ghosts (and in general, terms that are products of two combinations with opposite ghost number) and introduced the SK partner of the worldvolume metric. We remind the reader of notational convention \eqref{eq:snotation}, viz., 
\begin{equation}
\Ascr_a \equiv \As_a| \,, \qquad 
\tilde{\gref}_{ab} \equiv \Dsf_\theta  \Dsf_{\bar{\theta}} \,\SF{\gref}_{ab} | \,, \qquad 
\tilde{\Ascr}_a\equiv \Dsf_{\theta} \SF{\mathscr F}_{\thetab a}| \,, \qquad 
\etaref^{(ab)(cd)} \equiv \SF{\etaref}^{(ab)(cd)}| \,.
\label{eq:notegs}
\end{equation}	
This is also structure seen by \cite{Kovtun:2014hpa} who derived it via the MSR construction \cite{Martin:1973zz} adopted to relativistic fluid dynamics.\footnote{It was an oversight on our part to not draw this connection in the previous versions of this article. We thank Hong Liu for alerting us to this fact.}  It is thus very satisfying that a Wilson-type symmetry based construction of the action for an effective field theory, as given above, naturally reproduces the answer from an MSR like construction in \cite{Kovtun:2014hpa} which assumes the equations of motion beforehand.

The hydrodynamical stress tensor is obtained by varying with respect to the difference source $\shref_{ab}$ introduced in \eqref{eq:grefh}. Thus we define the hydrodynamic stress tensor
\begin{equation}
\TEMref^{ab}_\text{wv} = \frac{2(1+\Kref^e{\Ascr}_e)}{\sqrt{-\gref}}\, \frac{\delta \Lagref_\text{wv}}{\delta\shref_{ab}}\,.
\label{eq:Twvdef}
\end{equation}	
We will however adapt a trick to obtain the desired answer in a more straightforward manner.  First  let us  introduce a Hubbard-Stratonovich noise field $\noiseg_{ab}$ to rewrite \eqref{eq:classLTa} in a more suggestive form as
\begin{align}
\Lag_\text{wv} &=  \frac{\sqrt{-\gref}}{1+\Kref^e \Ascr_e} \;\bigg\{
	\frac{1}{2}\left[\TEMref_{\Lagref}^{ab}  - \frac{i}{2} \etaref^{(ab)(cd)}\, \left(  {\mathscr F}_{\theta\bar{\theta}}, 
	\gref_{cd}\right)_{\Kref}-\half \left(\etaref^{(ab)(cd)} + \etaref^{(cd)(ab)} \right) \noiseg_{cd}  
	\right] \tilde{\gref}_{ab}
 \nonumber \\
 & \qquad
	 + \frac{i}{8}\, \left(\etaref^{(ab)(cd)} + \etaref^{(cd)(ab)} \right) \noiseg_{ab} \noiseg_{cd}
	 -\Nref_{\Lagref}^a \;\tilde{\Ascr}_a
	+ \ldots \bigg\}\,.
\label{eq:classLT}
\end{align}
The first line above along with the last term in the second line, we then recognize as comprising Class $\LT$ Lagrangian postulated in \cite{Haehl:2014zda,Haehl:2015pja}. The extension of the adiabatic Lagrangian to include dissipative terms naturally comes with the noise field $\noiseg_{ab}$.   The second line gives a Gaussian measure with zero mean and an Avogadro suppressed variance  of ${\mathscr O}\left(\eta^{-1}\right)$ to various realizations of the noise field $\noiseg_{ab}$. To wit, the noise field is taken to be a random variate drawn from an ensemble with
\begin{equation}
\vev{\noiseg_{ab}} =  0 \,, \qquad \vev{\noiseg_{ab} \noiseg_{cd}} \sim \left(\etaref^{-1}\right)_{abcd} \propto \frac{1}{\eta}\,.
\label{}
\end{equation}	
Note that for this interpretation to work (and for the path integral to converge), the symmetric part of $\etaref$-tensor should be a positive operator acting on the space of symmetric 2-tensors. This will later turn out to be the tensor that controls dissipation and its positive definiteness will translate into the statement of the second law. That the same tensor controls the width of the noise is natural from the point of view of the fluctuation-dissipation theorem.

The full fluid dynamical stress tensor can now be read off directly by varying with respect to $\tilde{\gref}_{ab}$ itself  using the earlier results from \cite{Haehl:2015pja}\footnote{ We are skirting a technical subtlety here: $\tilde{g}_{\mu\nu}$ used in \cite{Haehl:2015pja} is not the push-forward of $\tilde{\gref}_{ab}$ here. The variables used here are more naturally adapted to the SK average and difference fields, so would correspond to linear combinations of the pull-backs of $g^\skR_{\mu\nu}$ and $g^\skL_{\mu\nu}$ introduced in equation (15.1) of that work. The difference matters not only from the additional $\UT$ gauge field pieces, but  also for the noise contributions. We are quoting here the physical average stress tensor in $\TEMref^{ab}_\text{wv}$, while the Class $\LT$ stress tensor of \cite{Haehl:2015pja} is for $T^{\mu\nu}_\skR$. The distinction is inconsequential for non-noise terms, but one needs to be careful with the identifications to obtain the noise contribution correctly.}
\begin{equation}
\begin{split}
 {\TEMref}_\text{wv}^{ab} &\equiv \frac{2(1+\Kref^e{\Ascr}_e)}{\sqrt{-{\gref}}} \frac{\delta {\Lagref}_\text{wv}}{\delta \tilde{\gref}_{ab}} \bigg|_{\tilde{\gref}_{ab} = 0}\\
 &  = \TEMref_{\Lagref}^{ab}  - \frac{i}{2} \etaref^{(ab)(cd)}\, \left(  {\mathscr F}_{\theta\bar{\theta}}, 
  	\gref_{cd}\right)_{\Kref}-\half \left(\etaref^{(ab)(cd)} + \etaref^{(cd)(ab)} \right) \noiseg_{cd}\,.
\end{split}
\label{eq:Twv}
\end{equation}
Varying $\tilde{X}^\mu$ inside $\tilde{\gref}_{ab}$ then yields the conservation law for ${\TEMref}_\text{wv}^{ab}$ as the equations of motion.

Finally, the $\UT$ gauge field $\SF{\Ascr}_a$ couples to the (negative of the) correct entropy current. To see this, 
we first observe that once $\tilde{\gref}_{ab}$ is written in terms of the elementary fields, $\tilde{\Ascr}_a$  appears in it
as $\tilde{\gref}_{ab}\ni \Kref_a \tilde{\Ascr}_b+ \Kref_b \tilde{\Ascr}_a + \,{\mathscr O}(\Ascr^2)$. This ensures that the $\tilde{\gref}_{ab}$ part of 
the action couples to the negative of the canonical entropy current with $\tilde{\Ascr}_a$ via a term of the form 
$ \Kref_a \,\TEMref^{ab}_\text{wv}\,\tilde{\Ascr}_b $. It remains to show that $\Nref_{\Lagref}^a$ gives the correct non-canonical
part. In particular, we want $\Nref^a=-\frac{\Gref^a}{\Tref}$ with $\Gref^a$ being the grand canonical free-energy
current and $\Tref=\prn{-\gref_{ab}\,\Kref^a \Kref^b}^{-1/2}$ being the temperature field.

The contributions to $\Nref_{\Lagref}^a$ come from the measure and the covariant derivatives 
$\mathcal{D}_a$ acting on $\gref_{ab}$ inside $\Lagref$ . When $\SF{\mathscr{A}}_e=0$,  $\Nref_{\Lagref}^a$ is the Noether current associated with thermal translations which is the correct non-canonical part of the entropy current. This coupling is  precisely what is expected from the Class $\LT$ construction of \cite{Haehl:2015pja}.
Note that there are  naturally ghost  contributions to the free energy (and hence entropy) current. This ensures that the full entropy current being a conserved symmetry current does not contradict the fact that its bosonic part satisfies a second law inequality. 

In order to get dissipation, the CPT ${\mathbb Z}_2$ should be broken spontaneously. We claim that the order parameter of this spontaneous symmetry breaking is the expectation value of ${\mathscr F}_{\theta\thetab}$. Thus, the dissipative fluid is a phase with $ \vev{\Fscr_{\theta\thetab}}\neq 0$ and for convenience, we will choose   $i \, \vev{\Fscr_{\theta\thetab}} = 1$.\footnote{ Note that  we have taken the $\UT$ covariant derivatives to be $\partial_I + (\As_I, \,\cdot\,)_\Kref$ where the $\UT$ generator is anti-hermitian. Thus expectation values of $\As_I$ or  $\SF{\Fscr}_{IJ}$ have to be purely imaginary.} With this understanding, the Class $\LT$ Lagrangian has been extended to include dissipative terms.
The second line we claim has to do with the fluctuations engendered by such dissipation.  At this stage we can immediately see that the correct constitutive relations are obtained 
for we can now port the results of Class $\LT$ of these works.
To illustrate this abstract formalism, let us compute two explicit examples of interest.
 
\paragraph{Viscous neutral fluids (first order):} 
First order neutral fluids are described by a Landau-Ginzburg type pressure and two Class D coefficients, viz., shear and bulk viscosities $\eta, \zeta$. The effective action describing these three kinds of transport is parametrized by plugging the following data into \eqref{eq:classLTa} (or equivalently into \eqref{eq:petalag})
\begin{align}
 \Lagref_{(1)} = p(\Tref)\,, \qquad \qquad \etaref^{(ab)(cd)}_{(1)} &=   \zeta(\Tref) \, \Tref \, \Pref ^{ab}\, \Pref^{cd} + 2  \, \eta(\Tref ) \, \Tref\, \Pref ^{c\langle a} \,\Pref ^{b\rangle d}  \,.
 \label{eq:neutralv}
\end{align}
We use the standard conventions for the fluid tensors on the worldvolume. To wit, 
\begin{equation}
\begin{split}
&\Pref_{ab} = \gref_{ab} + \uref_a\, \uref_b \,, \qquad
\Tref=\prn{-\gref_{ab}\,\Kref^a \Kref^b}^{-1/2}  \,, \qquad
\uref^a = \Tref\, \Kref^a\,,
 \\
&\nabla_a\, \uref_b = \sigref_{(ab)} + \omref_{[ab]} + \frac{1}{d-1}\,
\vartheta\, (\gref_{ab}+ \uref_a\,\uref_b) - \accref_b\, \uref_a\,,
 \\
& x^{\langle ab \rangle} = \Pref^a_{\ c}  \, \Pref^b_{\ d} \, x^{(cd)} - \frac{1}{d-1}\, \Pref^{ab}\,  \Pref_{cd}\, x^{cd} \,.
\end{split}
\label{eq:fluiddata}
\end{equation}	
Plugging this into \eqref{eq:petalag}, we can do the relevant superspace integral to obtain \eqref{eq:classLTa}. This  when varied with respect to the difference metric leads to the familiar viscous fluid hydrodynamic stress tensor
\begin{align}
\TEMref^{ab}_{(1)} &=  \epsilon(\Tref)\, \uref^a \,\uref^b + p(\Tref) \, \Pref^{ab} - \zeta(\Tref) \, \vartheta \, \Pref^{ab}
-2 \, \eta(\Tref) \, \sigref^{ab} \nonumber \\
 & \qquad -\; 
 	 \Tref\, \prn{  \,  \zeta(\Tref) \, \Pref ^{ab}\, \Pref^{cd}\, \noiseg_{cd} 
 	 + 2\,   \, \eta(\Tref ) \, \noiseg^{\langle ab\rangle}}\, , 
\label{eq:Tone}
\end{align}
where we identified the energy density $\epsilon(\Tref) = \Tref \frac{dp}{d\Tref} -p$. The second line   can be interpreted as the stochastic noise contribution to the stress tensor which is consistent with the linearized expressions obtained in \cite{Kovtun:2014hpa}.  
Note that we also obtain the correct free energy current from the variation with respect to the $\UT$ gauge field 
\begin{align}
 \Gref^e_{(1)}\big{|}_\text{bosonic} = - p(\Tref)\, \uref^e\,,
\end{align}
where we dropped the ghost contributions for simplicity.
 
The worldvolume stress tensor \eqref{eq:Tone} can be pushed-forward to the physical target space using $\partial_a X^\mu$ to obtain the fluid dynamical energy-momentum $T^{\mu\nu}$ (which will take exactly the same form).

\paragraph{Conformal neutral fluid  (up to second order):} The second example we turn to is the well understood case of conformal fluids. The zeroth and first order data is as in \eqref{eq:neutralv}  with the tracelessness of the stress tensor following from scale invariance demanding that $\zeta =0$. At second order there are 5 transport coefficients \cite{Baier:2007ix,Bhattacharyya:2008jc}: 3 in Class L, 1 in Class B\footnote{ We remind the reader that Class B terms include more familiar transport such as Hall viscosity and conductivity in three dimensional parity-violating systems. Their presence in parity-even fluids has not been fully explored.} and 1 in the dissipative Class D, using the basis of \cite{Haehl:2015pja}. We claim that the non Class B second order constitutive relations  follow from the  superspace Lagrangian:\footnote{ The  fifth Class B transport coefficient usually denoted $\lambda_2$,  contributes to the constitutive relations \eqref{eq:conffluid} as 
$\TEMref^{ab}_{\text{(2,B)}} =  \left(\lambda_2 + 2\,\tau -2 \kappa\right)\, \sigref^{\langle ac} \omref_{c}^{\ b\rangle} $. Naively, an intertwiner   \cite{Haehl:2015pja}
$$\etaref^{(ab)(cd)}_\text{(2,B)} = -(\lambda_2 + 2 \tau - 2 \kappa) \, \frac{\Tref}{2}\, \left(\omref^{c\langle a} \Pref^{b\rangle d} - \left[(ab) \leftrightarrow (cd) \right] \right) ,$$
 which is antisymmetric under the pairwise index exchange will give such a contribution, but this term is forbidden by the CPT invariance \eqref{eq:etacpt}.  To be compatible with the generalized Onsager relation \eqref{eq:etacpt} we then need 
$\etaref^{(ab)(cd)}  = [ \etaref^{(cd)(ab)} ]^{\text{\sf{\tiny CPT}}} = - \etaref^{(cd)(ab)} $.
This requires an imaginary intertwiner that is in tension with the convergence of the path integral. 

A choice consistent  with CPT which does not destroy the convergence of the path integral can be obtained if we are willing to include a non-minimal coupling of the form
$$\etaref^{(ab)(cd)}_\text{(2,B)} = -i{\mathscr F}_{\theta\bar{\theta}} \,(\lambda_2 + 2 \tau - 2 \kappa) \, \frac{\Tref}{2}\, \left(\omref^{c\langle a} \Pref^{b\rangle d} - \left[(ab) \leftrightarrow (cd) \right] \right) \,.$$
This does give the required Class B term. We thank Amos Yarom for pointing this to us. 
  \label{fn:classB}}
\begin{align}
 \Lagref_{(2)}^\Wey &= - \frac{\kappa}{2(d-2)} \;  {}^\Wey\Rref + \frac{(\kappa-\tau)}{2} \sigref^{ab}\,\sigref_{ab} + \frac{(\lambda_3 - \kappa)}{4} \omref^{ab}\,\omref_{ab}  \,,\nonumber\\
 \etaref^{(ab)(cd)}_\text{(2,D)} &= -(\lambda_1-\kappa)\, \Tref  \left(\sigref^{c \langle a} \Pref^{b\rangle d} -\frac{1}{d-1} \sigref^{ab} \Pref^{cd} 
 \right) .
 \label{eq:Leta2}
\end{align}
In writing this expression we have introduced some notation specific to conformal fluids. As originally explained in
\cite{Loganayagam:2008is}  it is useful to work with a Weyl covariant connection 
${\sf W}_a$ that preserves  homogeneity under conformal rescaling.  
In particular, this leads to the following curvature tensors that appear here:
\begin{equation}
\begin{split}
{\sf W}_a &= \accref_a -\frac{\vartheta}{d-1}\, \uref_a\,, \\
{}^\Wey\Rref &= \Rref + 2(d-1)\left(\nabla_a {\sf W}^a - \frac{d-2}{2}\,  {\sf W}^2\right) .
\end{split}
\label{eq:weylcov}
\end{equation}	

Explicitly, varying  the action in \eqref{eq:petalag} using \eqref{eq:neutralv} and \eqref{eq:Leta2} 
leads to the conformal fluid energy-momentum tensor:
\begin{equation}
\begin{split}
{}^\Wey\TEMref^{ab} &= {}^\Wey\TEMref^{ab}_{(1)}  + {}^\Wey\TEMref^{ab}_{(2)} + {}^\Wey\TEMref^{ab}_\text{noise}
\\
{}^\Wey\TEMref^{ab}_{(1)} &=   \epsilon(\Tref)\, \uref^a \,\uref^b + p(\Tref) \, \Pref^{ab} -2 \, \eta(\Tref) \, \sigref^{ab} 
\\
{}^\Wey\TEMref^{ab}_{(2)} &=  (\lambda_1 - \kappa)\, \sigref^{\langle ac} \sigref^{b\rangle}_{\,c}
+ \tau \left( \uref^c \DWeyl_c \sigref^{ab} - 2\, \sigref^{\langle ac} \omref_{ c}^{\;b\rangle}\right)  
\\
& \qquad +\;  
\lambda_3 \, \omref^{\langle ac}\omref_{ c}^{\ b\rangle } + 
\kappa \,\left(\Cref^{acbd}\,\uref_ c\,\uref_d + \sigref^{\langle ac}\,\sigref_{ c}^{b\rangle} + 2\, \sigref^{\langle ac} 
\omref_{ c}^{\ b\rangle} \right) \,,
\\
{}^\Wey\TEMref^{ab}_\text{noise} &= -2\,\eta\,\Tref \, \ \noiseg^{\langle ab\rangle} + (\lambda_1-\kappa)\,\Tref \left( \sigref^{c \langle a} \noiseg ^{b\rangle}_{\ c} -\frac{1}{d-1} \, \sigref^{ab} \, \Pref^{cd} \, \noiseg_{cd}\right) .
\end{split}
\label{eq:conffluid}
\end{equation}
The free energy current also takes the expected form; the  bosonic part which is a bit involved can be read off from \cite{Haehl:2015pja} (Appendix F).  

As written the energy-momentum tensor \eqref{eq:conffluid}  includes three distinct classes of transport in the eightfold classification of \cite{Haehl:2015pja}, which appear in a neutral conformal fluid. The pressure $p$ in $\Lagref_{(1)}$, the curvature ${}^\Wey\Rref$ and $\omref^2$ terms in $\Lagref_{(2)}$ are the hydrostatic terms (Class $\PS$) which were first discussed in \cite{Banerjee:2012iz,Jensen:2012jh}. The $\sigref^2$ term in $\Lagref_{(2)}$ is the Landau-Ginzburg Class $\LS$ term described in \cite{Haehl:2015pja}. These terms combine to form the Class L terms that are present in a neutral conformal fluid. If we restrict to hydrostatic equilibrium only the Class $\PS$ terms are allowed; everything else vanishes. The SK path integral constructed above localizes to the Euclidean path integral. The remaining terms involve the intertwining tensor: the contributions from $\etaref^{(ab)(cd)}_\text{(1)} $ and  $\etaref^{(ab)(cd)}_\text{(2,D)} $, which involve tensors symmetric under $(ab) \leftrightarrow (cd)$, are clearly the dissipative Class D terms. They are purely real, and thus consistent with the requirements of Onsager reciprocity as demanded by \eqref{eq:etacpt}.

It is instructive to adapt these results for a holographic conformal fluid for which the transport data is readily available from  \cite{Policastro:2001yc} and \cite{Baier:2007ix,Bhattacharyya:2008jc}.  For fluids dual to Einstein gravity, we have the aforementioned Lagrangian density parameterized by 
\begin{equation}
\begin{split}
p(\Tref) &= c_\text{eff}\, \left(\frac{4\pi \, \Tref}{d}\right)^d \,, \qquad 
\eta =   c_\text{eff}\, \left(\frac{4\pi \, \Tref}{d}\right)^{d-1} \,,\\  
\kappa & = \lambda_1 = 2\, c_\text{eff}\, \left(\frac{4\pi \, \Tref}{d}\right)^{d-2}\,, \qquad 
\lambda_2  = 2 (\kappa - \tau) \,, \qquad \lambda_3 =0  \,,\\
\tau &= 2\, c_\text{eff}\, \left(\frac{4\pi \, \Tref}{d}\right)^{d-2} 
	\left[ 1 + \frac{1}{d}\, \text{Harmonic}\left(\frac{2}{d}-1 \right) \right] ,
\end{split}
\label{}
\end{equation}
where $c_\text{eff} $ is the effective central charge of the QFT.\footnote{ For holographic theories it is convenient to normalize $c_\text{eff} = \frac{\ell_\text{AdS}^{d-1}}{16\pi\, G_N}$, so as to get a simple result for the Bekenstein-Hawking entropy. For $SU(N)$, ${\cal N}=4_\text{4d}$ SYM we obtain $c_\text{eff} = \frac{N^2}{8\pi^2}$.} 
This can be succinctly written as a superspace integral 
\begin{equation}
\begin{split}
\Lagref_\text{wv} = 
	c_\text{eff} \ &
	\int d\theta\, d\thetab \; \frac{\sqrt{-\SF{\gref}}}{1+\Kref^e\,\As_e} 
	 \bigg\{ \left(\frac{4\pi \,\SF{\Tref}}{d}\right)^{d} \left(1  - \frac{i\,d }{8\pi}\; \SF{\Pref}^{c\langle{a}} \SF{\Pref}^{b\rangle d} \ 
	 \Dsf_\theta \SF{\gref}_{ab} \; \Dsf_{\thetab} \SF{\gref}_{cd} \right)
	 \bigg.
 \\
& \bigg. \qquad \quad	 
	 - \; \left(\frac{4\pi \,\SF{\Tref}}{d}\right)^{d-2}  \bigg[ \frac{{}^\Wey\SF{\Rref}}{d-2} 
	+ \frac{1}{d} \,\text{Harmonic}\left(\frac{2}{d} -1\right) \SF{\sigref}^2 + 
	 \frac{1}{2} \SF{\omref}^2 \bigg] \bigg\}\,.
\end{split}
\label{}
\end{equation}
This expression generalizes the bosonic Class L Lagrangian given in \cite{Haehl:2014zda,Haehl:2015pja}  for the adiabatic part of the constitutive relations, cf.,  equation (14.35) of the latter.

 Holographic fluids described by Einstein-Hilbert gravity thus do not give the most general conformal fluid; as noticed in \cite{Haehl:2015pja} they miss out on the Class B term (owing to the $\lambda_2$ relation derived first in 
\cite{Haack:2008xx}) and pick  out the value of $\lambda_1$ that makes the second order dissipative contribution vanish. We have conjectured hitherto that this has to do with holographic fluids being optimal dissipators \cite{Haehl:2014zda}.

It is worth noting that the dissipative transport coefficients scale with the central charge $c_\text{eff}$. This means that the noise terms are suppressed by a factor of $c_\text{eff}^{-1}$. In familiar holographic systems the dissipative energy-momentum tensor has thus two contributions: a leading ${\mathscr O}(c_\text{eff})$ term from the non-noise contributions and a subleading ${\mathscr O}(1)$ term from the noise. This is indeed what one should expect in systems at large $N$ (or large central charge). The fluctuations which lead to macroscopic noise should be Avogadro suppressed by the total number of degrees of freedom.
%

\section{Fluctuation-Dissipation \& Jarzynski relation}
\label{sec:fd}

A central assumption in conventional formulation of hydrodynamics is the existence of an entropy current with non-negative divergence. Obtaining such a sign-definiteness constraint from an effective action, especially for a derived object such as entropy, a-priori appears difficult. However, one can use the underlying supersymmetry to obtain a Ward identity which leads immediately to the apposite convexity statement for entropy. This result is basically the statement of the Jarzynski relation \cite{Jarzynski:1997aa}
applied to hydrodynamics. 

To explain these statements we now turn to examining carefully the consequences of the spontaneous symmetry breaking that we postulated in  \S\ref{sec:nld} to get dissipation. Using the $\mathbb{Z}_2$ invariance of  $S_\text{wv}$ in \eqref{eq:petalag}, we now argue for an interesting relation for the fluctuations in the broken phase. We will necessarily be brief and  refer the reader to \cite{Gaspard:2012la} for a more detailed examination of the phenomenology of such a discrete symmetry breaking.  In fact, the situation here is analogous to spontaneous symmetry breaking in the Abelian-Higgs model, where if we gauge fix the phase of the (complex) Higgs field, then we end up with a St\"uckelberg mass term for the vector boson. The Ward identity for the broken abelian gauge transformation in this gauge is the London equation for the current. Our claim here  is that the corresponding statement for spontaneous CPT breaking in the presence of external sources is the Jarzynski fluctuation-dissipation relation.

To begin with, note that under a general $\UT$ gauge transformations, our hydrodynamic fields transform as 
\begin{equation}
\begin{split}
 \delta \SF{X}^\mu &= 
 	\SF{\Lambda}\, \lieD_\Kref \SF{X}^\mu =  \SF{\Lambda} \, \Kref^a \partial_a \SF{X}^\mu \,, 
 \\
 \delta \SF{\gref}_{ab} &= 
 	\SF{\Lambda}\,  \lieD_\Kref \,\SF{\gref}_{ab}  \,,
\\
\delta \SF{\Ascr}_a &= 
	(\SF{\Lambda},\SF{\Ascr}_a)_\Kref - \partial_a \SF{\Lambda} =  - \Dsf_a \SF{\Lambda}\,. 
\end{split}
\end{equation}
The key point we want to note here is that for super-gauge parameters that only have a top, i.e., $\thetab \theta$ component, the gauge transformation shifts  the top component of the  superfield by a Lie drag of its bottom component, i.e., for  $\SF{\Lambda}  = \thetab\,\theta\, \tilde{\Lambda}$ the transformations simplify to 
\begin{align}
 \delta \tilde{X}^\mu = \tilde{\Lambda} \, \Kref^a \partial_a {X}^\mu \,, \qquad 
 \delta \tilde{\gref}_{ab} = \tilde{\Lambda}\, \lieD_{\Kref} \; {\gref}_{ab} \,, \qquad
 \delta \tilde{\Ascr}_a &=- \mathcal{D}_a \tilde{\Lambda}\,.
 \label{eq:UTtop}
\end{align}

We now make the following claim:  CPT transformations are achieved by just performing particular $\UT$ gauge transformations.  Choosing to work in a particular gauge\footnote{ All statements below refer to a particular supergauge fixing; we use the Wess-Zumino gauge. Furthermore, we also set $\Ascr_a =0$ at the end of the variation (details will appear in \cite{Haehl:2015ab}).} the spontaneously broken CPT symmetry can be probed by $\SF{\Lambda}_\text{CPT} = - \thetab\theta\, \mathscr{F}_{\theta\thetab}$. Then from \eqref{eq:UTtop} we conclude that the transformation we seek shifts the fluid variables as:
\begin{align}
\tilde{\gref}_{ab} \mapsto \tilde{\gref}_{ab} - \big{(}  \mathscr{F}_{\theta\thetab}  , \, \gref_{ab} \big{)}_\Kref \,, \qquad
\tilde{\Ascr}_a \mapsto \tilde{\Ascr}_a +\mathcal{D}_a  \mathscr{F}_{\theta\thetab}  \,.
\end{align}
Implementing this we see that the worldvolume Lagrangian density transforms with an inhomogeneous piece (as expected due to the gauge fixing)
\begin{align}
  \Lag_\text{wv} \,&\mapsto\, \Lag_\text{wv} + \frac{\delta \Lag_\text{wv}}{\delta \tilde{\gref}_{ab}} \, \delta \tilde{\gref}_{ab}  + \frac{\delta \Lag_\text{wv}}{\delta \tilde{\Ascr}_{a}} \, \delta \tilde{\Ascr}_{a} \\
  &\;\;=\Lag_\text{wv}- \mathscr{F}_{\theta\thetab}  \left( \frac{1}{2} \, \TEMref_\text{wv}^{ab} \, \lieD_\Kref\, \gref_{ab} 
     - \mathcal{D}_a \Nref^a \right) + \text{boundary terms} \,.
\end{align}
The boundary terms above get contributions from the various integrations by parts performed in doing the transformations and superspace integrals. 

We now invoke the expectation value of $\vev{\Fscr_{\theta\bar{\theta}} }= -i$  and write the change in the action functional  $S_\text{wv} \equiv \int d^d\sigma \, \Lag_\text{wv} $ suggestively as
\begin{align}
 i\, S_\text{wv} \mapsto & \;  
     i \, S_\text{wv} - \Tref^{-1} \, (G_f-G_i +W )\,.
\label{eq:actionCPT}     
\end{align}
We introduced here the free energy difference and total work done by the external source:
\begin{align}
G_f - G_i &\equiv 
	-\Tref \, \int d^d \sigma \; \frac{\sqrt{-\gref}}{1+\Kref^e\, \Ascr_e} \; \mathcal{D}_a \Nref^a 
     =  	-\Tref
     \left[\int_{\Sigma_E} \; \Nref^a \, d^{d-1}S_a \right]\bigg{|}_{t_i}^{t_f}
 	\,,\\
W & \equiv 
 	  \Tref \int d^d\sigma \;\sqrt{-\gref}\,\left(\frac{1}{2} \, \TEMref_\text{wv}^{ab} \, \lieD_\Kref\, \gref_{ab} \right)\,,
\end{align}
with the assumption that the boundary terms will conspire to cancel out of this analysis.
The integral in $G_f-G_i$ over worldvolume time has been performed such as to localize onto a hydrostatic integral over the Euclidean base manifold $\Sigma_E$ with volume element $d^{d-1}S_a$ evaluated in the equilibrium configurations at initial and final times $t_i$, $t_f$. Note in particular that this integral is independent of the (generically non-adiabatic) protocol which takes the system from the initial to the final configuration and it includes contributions from the ghost superpartners of the fields. 

From equation \eqref{eq:actionCPT} we can  get the hydrodynamic fluctuation-dissipation result we seek following \cite{Mallick:2010su}.  The underlying topological symmetry implies the following Ward identity 
\begin{align}
\vev{e^{-\frac{W}{\Tref}}} = e^{ -\frac{1}{\Tref} \left( G_f - G_i \right)} ,
\label{eq:GenJarz}
\end{align}
i.e., the expectation value of the exponential of the work done is the exponential of the free energy difference. Using Jensen's inequality on the above we  obtain 
\begin{equation}
\vev{W} \geq G_f - G_i\,,
\label{eq:2law}
\end{equation}	
which asserts that entropy is produced in the system. In other words the generalized work relation \eqref{eq:GenJarz} implies the second law of thermodynamics, ensuring that our construction is consistent with the axioms of hydrodynamics.

\section{Discussion}
\label{sec:discuss}

Obtaining an effective action for dissipative hydrodynamics, as we have presaged in \S\ref{sec:intro}, is interesting not only for understanding the dynamics of quantum fields in generic non-equilibrium settings, but has also implications for other areas of physics. The advantage hydrodynamics has is that the phenomenological theory is very well understood and one has a clear target to attain to declare success. We have outlined here a series of general principles which enshrine the symmetries that such an effective field theory should respect. Crucial to the discussion are the topological symmetries inherent in the SK path integral construction and an emergent gauge symmetry which is the macroscopic manifestation of the KMS invariance of near-equilibrium systems. We have also identified a particular superspace component 
of this emergent field strength as the relevant order parameter for spontaneous symmetry breaking of CPT.

Equipped with the symmetries outlined in \S\ref{sec:fss} we have constructed a simple topological sigma model for the hydrodynamic modes. The construction naturally makes contact with the empirically obtained Class $\LT$ effective action in \cite{Haehl:2015pja} with the added feature of having various ghost contributions (which we have suppressed) to ensure that the theory has the correct topological symmetries. We have shown that this action reproduces correctly the known constitutive relations for neutral fluids.

Furthermore, the conserved current associated with the KMS gauge symmetry is the entropy current which satisfies the requirements of the second law of thermodynamics. Inspired by earlier work on non-equilibrium work relations \cite{Jarzynski:1997aa,Jarzynski:1997ab,Crooks:1998uq,Crooks:1999fk} we have argued that the natural way to view the second law is in terms of the Jarzynski equality. This has been argued to follow by carefully monitoring the invariances of the topological action under CPT \cite{Mallick:2010su,Gaspard:2013vl}  and our analysis for the hydrodynamic effective actions respects similar constraints. The hydrodynamic work relation implies that over the period of time that the fluid is driven out of equilibrium by an external source,  entropy is produced in the system.

The key hydrodynamic question we have refrained from addressing here is whether all known constitutive relations can be obtained from  our effective action. To claim that our construction is comprehensive, we should 
establish that the set of allowed terms in the topological sigma model is in one-to-one correspondence with the eightfold classification of hydrodynamic transport \cite{Haehl:2015pja}. We have seen clearly that three classes, viz., 
hydrostatic ($\PS$), hydrodynamic ($\LS$) and dissipative (D) are naturally encompassed in our formalism. These are the only ones encountered were we to study neutral conformal fluids which are holographically dual to large AdS black holes in Einstein-Hilbert gravity. 

The remaining five classes of transport involve three finite classes, viz., anomalous transport (A), hydrostatic vectors 
($\PV$), and  conserved entropy (C), in addition to two more infinite classes: Berry-like (B) and hydrodynamic vectors 
($\GV$). Of these, neutral fluids admit only Class B terms at second order; to see any of the other four classes we need to turn on flavour charges. As we have indicated at the end of \S\ref{sec:fss} it is not too difficult to  generalize our considerations to incorporate flavour symmetries. While there remain some details to be worked out, the fact that we land precisely on the Class $\LT$ Lagrangian, which we recall was cognizant of the seven adiabatic classes, lends us confidence that the topological sigma model will be compatible with the eightfold classification.  We will give a detailed account of how this works elsewhere.

Perhaps the most intriguing aspect of hydrodynamics is the lessons is holds for gravitational systems involving black holes. Via the fluid/gravity correspondence we know that the hydrodynamic description of holographic CFTs is given by the dynamics of large black holes in AdS --  the classical partial differential equations of  fluid dynamics are a subset of Einstein's equations. One should however expect  the correspondence to extend to include hydrodynamic fluctuations, which after all, go hand-in-hand with dissipation. For the hydrodynamic theory, getting the noise terms in \eqref{eq:classLT} correctly is contingent on the topological symmetries. While we have elided in our presentation over the ghost fields contained in the supermultiplets for simplicity, it is their presence that ensures the correct coupling between the noise and the dissipative terms. Moreover, in our formalism, while we recover the familiar entropy current by a Noether construction analogous to the story for black holes, the discussion in the text should make it clear that the current that is conserved includes non-trivial ghost contributions. This allows the bosonic part of the entropy to be non-conserved and produced consistent with the second law, provided the ghost entropy flux compensates for the same.

We suggest that herein lies a crucial moral that ought to extend to black hole physics. The origin of dissipation in the classical theory stems from field modes disappearing behind the black hole horizon. Since it is this dissipation that leads to entropy growth, one may naturally link this statement with the ghost degrees of freedom becoming more relevant at the thermal scale. This picture is consistent with the breakdown of the scale/radius duality on horizon scales, and furthermore suggests that any description of the interior of the black hole should involve understanding of these ghosts more throughly. As a natural corollary one may want to view the black hole interior as comprised of a ghost condensate of sorts.
Further, the statement that $ \vev{\Fscr_{\theta\thetab}}\neq 0$ on the fluid dynamical side should have a corresponding counterpart in the 
gravity description and understanding its phenomenology is likely to be crucial in characterizing the  apparent non-unitarity in black hole
physics. Heuristically, this picture suggests potential resolutions to various puzzles encountered in the subject over the years, but to make clear statements, one needs to construct the topological supergravity theory, i.e., the closed string dual of the hydrodynamic theory.  This theory in particular has to reproduce the noise contribution of  \eqref{eq:conffluid}. We leave this fascinating issue as an interesting challenge for our formalism to reproduce in the future.

\acknowledgments

It is a pleasure to thank V.~Hubeny, H.~Liu, J.~Maldacena, S.~Minwalla for discussions. FH would like to thank Perimeter Institute for hospitality while this work was finalized.  RL would  like to thank the organizers and speakers of the ICTS workshop on non-equilibrium statistical physics for enlightening discussions on large deviation theory.  FH and RL would also like to thank Center for Quantum Mathematics and Physics (QMAP) at UC Davis for hospitality during the course of the project.

FH is supported by a Durham Doctoral Fellowship and by a Visiting Graduate Fellowship of Perimeter Institute. Research at Perimeter Institute is supported by the Government of Canada through Industry Canada and by the Province of Ontario through the Ministry of Research and Innovation. RL gratefully acknowledges support from International Centre for Theoretical Sciences (ICTS), Tata institute of fundamental 
research, Bengaluru. RL would also like to acknowledge his debt to all those who have generously supported and encouraged the pursuit of science in India. MR was supported in part by the ERC Consolidator Grant Agreement ERC-2013-CoG-615443: SPiN and by the FQXi  grant ``Measures of Holographic Information"  (FQXi-RFP3-1334).


\providecommand{\href}[2]{#2}\begingroup\raggedright\endgroup

\end{document}